# Ferromagnetic resonance in submicron permalloy stripes

E. V. Skorohodov*[1,2], R. V. Gorev[1], R. R. Yakubov[2], E. S. Demidov[2], Yu. V. Khivintsev[3], Yu. A. Filimonov[3], and V. L. Mironov[1,2*]

[1]Institute for Physics of Microstructures RAS, 603950, Nyzhny Novgorod, Russia

[2]Lobachevsky State University, 603950, Nyzhny Novgorod, Russia

[3]Kotel'nikov Institute of Radio-engineering and Electronics RAS, Saratov branch, 410019, Saratov, Russia

*Corresponding author: evgeny@ipmras.ru

**Abstract**

We present the results of systematic experimental investigations and micromagnetic simulations for the ferromagnetic resonance in rectangular permalloy microstripes. It is shown that the resonant magnetization oscillations have a complex spatial structure including a quasi-homogeneous precession, lateral spin-wave resonances and localized edge modes, which strongly depend on sample orientation in an external magnetic field.

**1. Introduction**

The microwave properties of the ferromagnetic planar patterned structures (PPS) is the subject of the recent intensive investigations motivated by perspective applications in magnetoelectronics, spintronics and data processing [1,2]. In particular, the PPS consisting of ferromagnetic elements with different shape and spatial arrangement are considered as the effective tuned filters for the ultra high frequency (UHF) electromagnetic radiation [3,4]. The ferromagnetic resonance (FMR) and spin wave resonances (SWR) in PPS strongly depend on the internal fields connected with the shape anisotropy, exchange and magnetostatic interaction. As a result the absorption spectrum of PPS is defined by the different geometric factors such as shape, size and spatial arrangement of elements in PPS. This opens the wide opportunities for tuning of UHF absorption by changing the architecture of PPS using the nanolithography methods. In this regard, the special researches are focused on the study of ferromagnetic resonance in rectangular microstripes, considered as one of the main structural elements of planar UHF microsystems. Due to the high shape anisotropy the microstripes have a uniform magnetic state and can be used in the microwave devices without external magnetizing. In particular, the PPS with strong magnetostatic interaction enables the realization of tuned filters with different absorption spectra, which can be switched by the external magnetic field [5]. Partially the investigations of the magnetization oscillations spectra depending on the size and shape of microstripes were reported in [6-11]. The special attention was paid to the analysis of uniform precession and localized edge modes arising from the non-homogeneity of internal magnetostatic field near the boundaries of microstrip [9-12]. However, the systematic FMR studies for different microstrip orientations in an external magnetic field have not been performed.

The dynamic properties of PPS systems are widely studied by micromagnetic modeling based on the numerical solution of the Landau-Lifshitz equation. In general, authors simulate the spectra of microstrip eigenfrequencies based on Fourier analysis of relaxation oscillations of the magnetization [12]. However, experimentally the FMR is studied often by the analysis of the microwave absorption at a given frequency in dependence on the magnitude and direction of sweeping external dc magnetic field. Appropriate modeling of the spectra and spatial distributions of steady-state magnetization oscillations in external magnetic fields was not discussed.

In current paper we present the results of experimental measurements and micromagnetic simulations of magnetization given to the analysis of mode structure and spatial distribution of spin oscillation oscillations in planar rectangular microstrip for different sample orientations in an external magnetic field. The main attention is observe the spin-wave resonances (the formation of standing spin waves).

## 2. Experiments and methods

The array of permalloy ($Ni_{80}Fe_{20}$) rectangular stripes $3000{\times}500{\times}30$ $nm^3$ (the separation between stripe's centers is about 6 µm) were fabricated by electron-beam lithography and lift-off process. At the first stage the 150 nm resist was deposited on Si substrate by centrifuging. Then initial mask in the form of rectangular stripes array was formed in negative electron resist using scanning electron microscope SUPRA 50VP with lithographic facility ELPHY PLUS. After that the irradiated areas in resist were removed in a selective organic solvent. At the next step the sample was covered by permalloy layer (30 nm thick) using magnetron sputtering. At the final stage the areas of permalloy layer except the strip array was removed in the lift-off process. The typical SEM image of strip array is presented in Fig. 1(a).

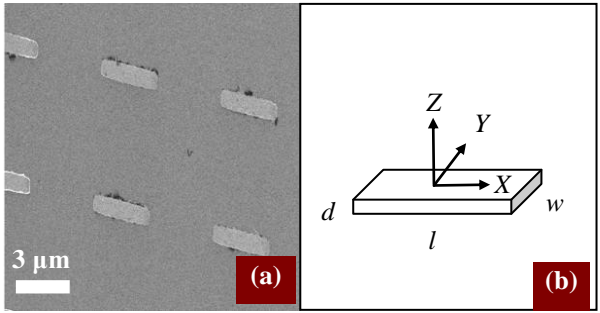

**Fig. 1.** (a) The SEM image of permalloy rectangular stripes $3000{\times}500{\times}30$ $nm^3$. The separation between stripes is 3 µm. (b) The associated coordinate system, connected with the sample.

The selected distance between the microstrips realizes the weak magnetostatic interaction between neighboring elements in the array, so that the sample FMR spectrum corresponds to a FMR spectrum of separate microstrip. Note that all microstrip dimensions are less than the path length of spin waves in permalloy that allows us to

The FMR measurements were performed with Bruker EMX Plus-10/12 electron paramagnetic resonance spectrometer using a $TE_{011}$ mode of cylindrical resonant cavity and dc magnetic field ($H$) up to 2 T. The polarized microwave magnetic field $\vec{\tilde{h}}$ with frequency 9.8 GHz was perpendicular to the $H$. The samples were driven through the resonance by sweeping the $H$.

Theoretically, the microwave absorption in microstrip was studied by computer micromagnetic modeling based on numerical solution of Landau-Lifshitz-Gilbert equation using the standard object oriented micromagnetic framework (OOMMF) code [13]. We investigated the dependencies of microvawe power absorption on the magnitude of swept $H$. Algorithm of numerical simulation includes follow stages. 1) The external magnetic field is applied. 2) The standing spin waves is excited by microwave magnetic field. 3) The dependence of averaged variable magnetization amplitude from the magnitude of external magnetic field is graphed. This dependence is a curve microvawe power absorption on the magnitude of swept $H$ because absorbed power is proportional to the sample averaged variable magnetization amplitude. In addition, to visualize the mode composition of oscillations we calculated the time dependencies of spatial distributions for different magnetization components. The calculations were performed for the typical permalloy parameters: the saturation magnetization was $M_S = 8 \times 10^5$ A/m, the exchange stiffness was A = $8.4 \times 10^{-12}$ J/m, and the damping constant was 0.01 (see [14]). We omitted magnetocrystalline anisotropy, assuming a polycrystalline structure of our samples, the perpendicular uniaxial anisotropy $K = 3.9 \times 10^4$ $J/m^3$. The microwave frequency was 9.8 GHz. In calculations the microstrip $3000{\times}500{\times}30$ $nm^3$ was discretized into rectangular parallelepipeds with a square base of size δ = 10 nm in the *x,y* plane and height *h* = 30 nm. The choice of geometrical dimensions and aspect ratio (the ratio of thickness, width and length) of microstripes was motivated by convenience for the experimental and theoretical study of the eigenmodes of magnetization oscillations. In case of high aspect ratio the magnetostatic interaction would give rise to noticeable shape anisotropy and substantial inhomogeneity of internal magnetic field that allows studying the influence of these effects on FMR spectrum.

## 3. Results and discussion

We investigated the FMR response in dependence on sample orientation relatively $\vec{H}$. In all experiments the exciting magnetic field $\vec{h}$ was directed in the sample plane as the oscillation excitation efficiency in this case is maximal. At the first step we investigated the FMR in geometry when the external magnetic field was applied along the long side of microstripe (X axis in fig. 1b). Fig. 2a shows experimental (dashed red curve) and numerically calculated (solid blue curve) FMR spectrum for this geometry. It is seen that there is a good agreement between the experiment and simulation. As is seen from fig. 2b the resonant oscillations corresponding to resonance fields 1 - 3 have complex spatial distributions, which are superposition of different modes of longitudinal spin-wave resonance along the microstrip. It is seen that there is a superposition of long-wave and short-wave modes. The long-wave mode is defined by magnetodipole interaction, short-wave mode is defined by exchange interaction.

To explain the observed mode composition we calculated the spectra of spin waves for the microstripe in an external magnetic field directed along the X axis. The theory of the dipole-exchange wave's spectra was developed in [15-17]. Following the proposed approach we assumed that the magnetic moments on the surfaces are not fixed and the lower modes do not depend on the *z* coordinate. It is known that the demagnetizing field of a rectangular parallelepiped is substantially nonuniform that leads to the complicated dependence of the variable component magnetization on coordinates *x*, *y*. However, approximately the distribution of transverse component of the variable magnetization can be written as follows [17]:

$$m_{y,z}(x,y) \approx M_s \cos(k_{xn}x)\cos(k_{ym}y), \quad (1)$$

where $M_s$ is magnetization in saturation; $k_{xn}=\pi(n+1)/l$ and $k_{ym}=\pi(m+1)/w$ ; *l* and *w* are the length and width of the microstrip respectively. The deviation of the variable component of the magnetization from the cosine function is $d/l$ (where *d* is the microstrip thickness). On the other hand the amendment to the spectrum is $(d/l)^2$, which is not substantial for the stripes with high aspect ratio. Thus, the spectrum of magnetization oscillations can be represented as follows [16,17]:

$$\omega_{nm}^2 = \gamma M_s (\omega_H^{nm} + J\omega_M k_{nm}^2)(\omega_H^{nm} + J\omega_M k_{nm}^2 + \omega_M F_{mn}), \quad (2)$$

where γ is gyromagnetic ratio; $\omega_M = 4\pi M_s$; $\omega_H^{nm} = \gamma H - \omega_M N_{nm}$; $N_{nm}$ are demagnetizing factors for different modes in the microstrip; $F_{mn}$ are the matrix element of the dipole-dipole interaction; *J* is the exchange stiffness constant. Using (2) and taking into account the frequency of the microwave oscillations $\nu_0 = 9.8$ GHz, it is possible to find the resonant magnetic fields for each FMR mode. As can be seen from fig. 2b, in the resonances we observe a superposition of two modes with low and high spatial frequencies (wave numbers). It is connected with the excitation of both dipole spin waves with small $k_{xn}$ as well as exchange waves with large $k_{xn}$. The dispersion relation in this case is non-monotonic and there is the degeneracy in the spectrum.

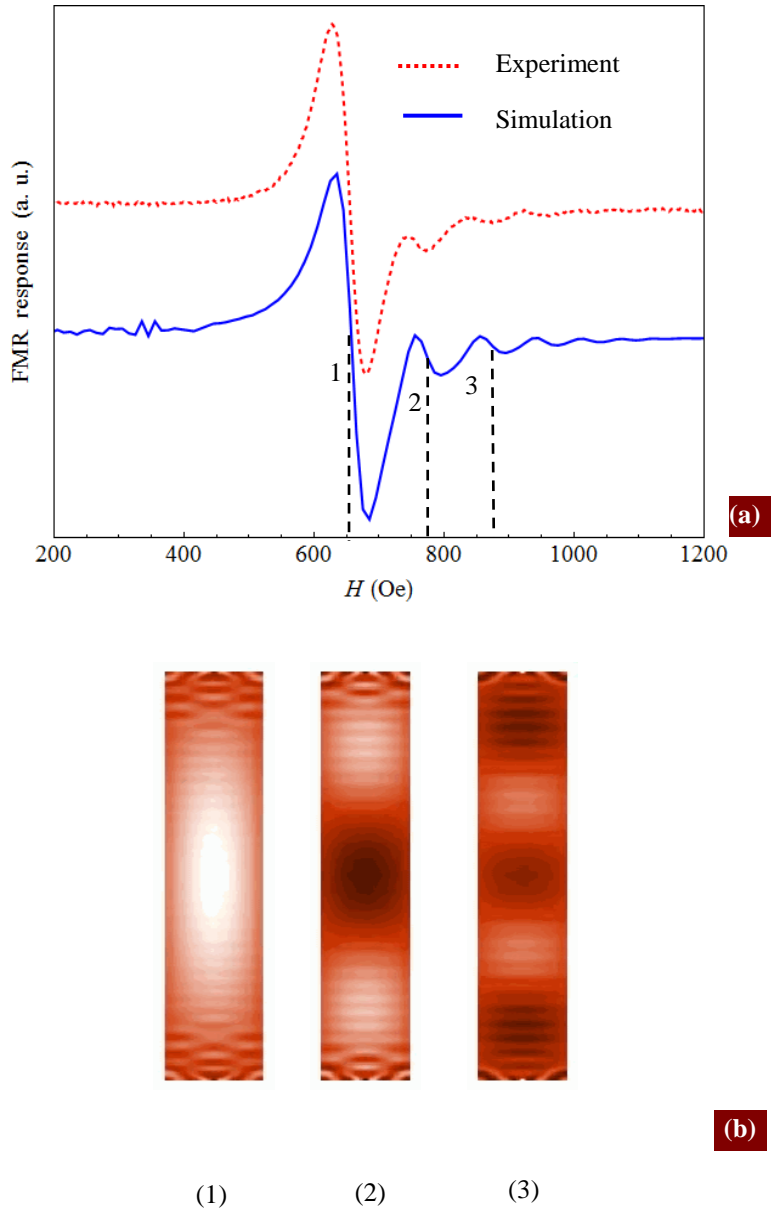

**Fig. 2.** (a) The experimental (dashed red curve) and simulated (solid blue curve) microstripe FMR spectra (external magnetic field applied along X axis). (b) The spatial distributions of the magnetization oscillations (amplitude of the z-component) in different resonant fields. (1) is for $H = 660$ Oe; (2) is for $H = 770$ Oe; (3) is for $H = 870$ Oe.

Calculations shows that the resonances are realized at $H = 645$ Oe for the modes with $n = 0$ and $n = 84$ (resonance No 1); at $H = 740$ Oe for the modes with $n = 2$ and $n = 82$ (resonance No 2); at $H = 850$ Oe for the modes with $n = 4$ and $n = 78$ (resonance No 3) respectively. For this resonance spin waves the mode number $m=0$. These values of the resonance fields and numbers of longitudinal SWR modes are in good agreement with the results of micromagnetic simulations (fig. 2a).

The edge roughness has a big influence for configuration when external magnetic field applied along short side of sample (Y axis in fig 1b). At the first we considered microstripe without edge roughness. As is seen from fig 3a there is a good agreement between the experiment and the simulation. But intensities of edge modes are different. For this experimental configuration analytical calculation was performed also. The degeneracy of resonance magnetic field from mode number is observed for modes with wave numbers ($n = 2$; $m = 0$) and ($n = 2$; $m = 12$) for resonance No 1; for modes with wave numbers ($n = 0$; $m = 0$) and ($n = 0$; $m = 10$) for resonance No 2; for modes with wave numbers ($n = 4$; $m = 4$) and ($n = 4$; $m = 8$) for resonance No 3. For other modes degeneracy of resonance magnetic field from mode number is not observed. The resonances No 4 and No 5 correspond to edge modes. The edge modes corresponding resonances No 4 and No 5 have different numbers of half-waves. The resonance No 4 has three half-waves, resonance No 5 has one half-wave (fig. 3b).

To decrease intensity of edge mode we take into account edge roughness with standard deviation 15 nm. In this case edge modes are suppressed (fig. 4a). The visualization of the spatial distributions of the variable magnetization (fig. 4b) shows that edge roughness does weak changing of mode structure.

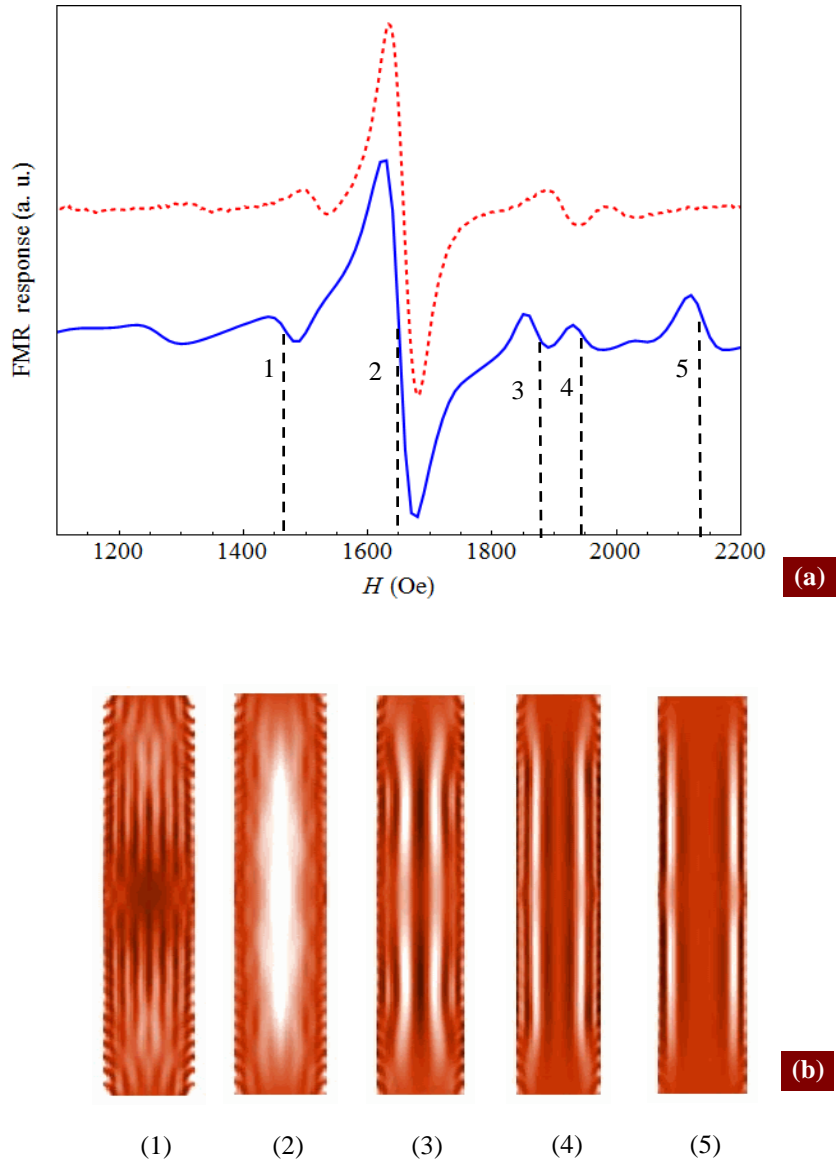

**Fig. 3.** (a) The experimental (dashed red curve) and simulated (solid blue curve) microstripe FMR spectra (external magnetic field applied along Y axis). (b) the spatial distributions of the magnetization oscillations (amplitude of the z-component) in different resonant fields. (1) is for $H = 1493$ Oe; (2) is for $H = 1650$ Oe; (3) is for $H = 1870$ Oe; (4) is for $H = 1940$ Oe; (5) is for $H = 2140$ Oe.

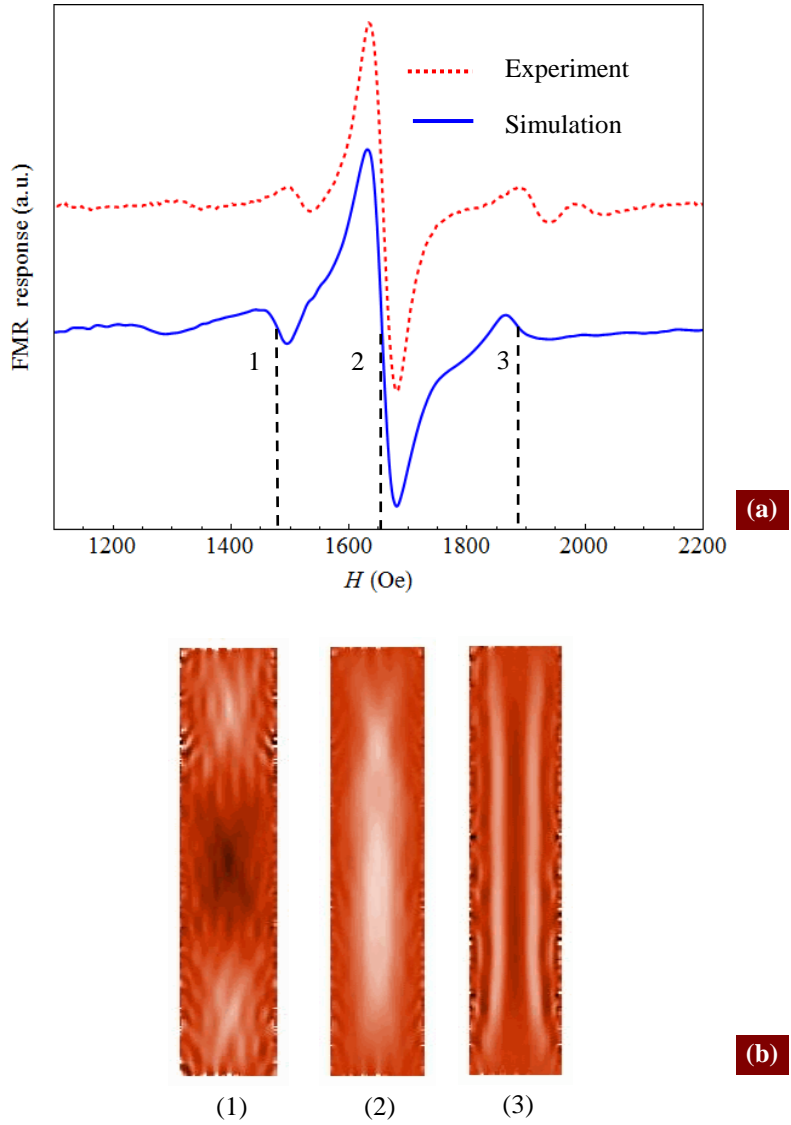

**Fig. 4** (a) The experimental (dashed red curve) and simulated (solid blue curve) microstripe FMR spectra for (external magnetic field applied along Y axis). (b) the spatial distributions of the magnetization oscillations (amplitude of the z-component) in different resonant fields. (1) is for $H = 1493$ Oe; (2) is for $H = 1650$ Oe; (3) is for $H = 1870$ Oe

Fig 5a shows experimental and simulated microstripe FMR spectra when external magnetic field applied perpendicular to sample plane (Z axis on fig 1(b)). For this geometry main resonance corresponding main mode and spin-wave resonances locating left from main resonance is observed. Also visualization of variable magnetization spatial distribution was made. For resonances No 1-3 the amplitude of magnetization oscillation is changed along a long side only, while for the resonances No 4 and No 5 the amplitudes of magnetization oscillation is changed along long and short sides. In this geometry degeneracy of resonance magnetic field from mode number is absent but the superposition of long- and short-wave modes is observed. For $H = 10850$ Oe resonance modes are ($n=0$; $m=2$) and ($n=18$; $m=0$). For $H = 10320$ Oe resonance modes are ($n=0$; $m=4$) and ($n=34$; $m=0$).

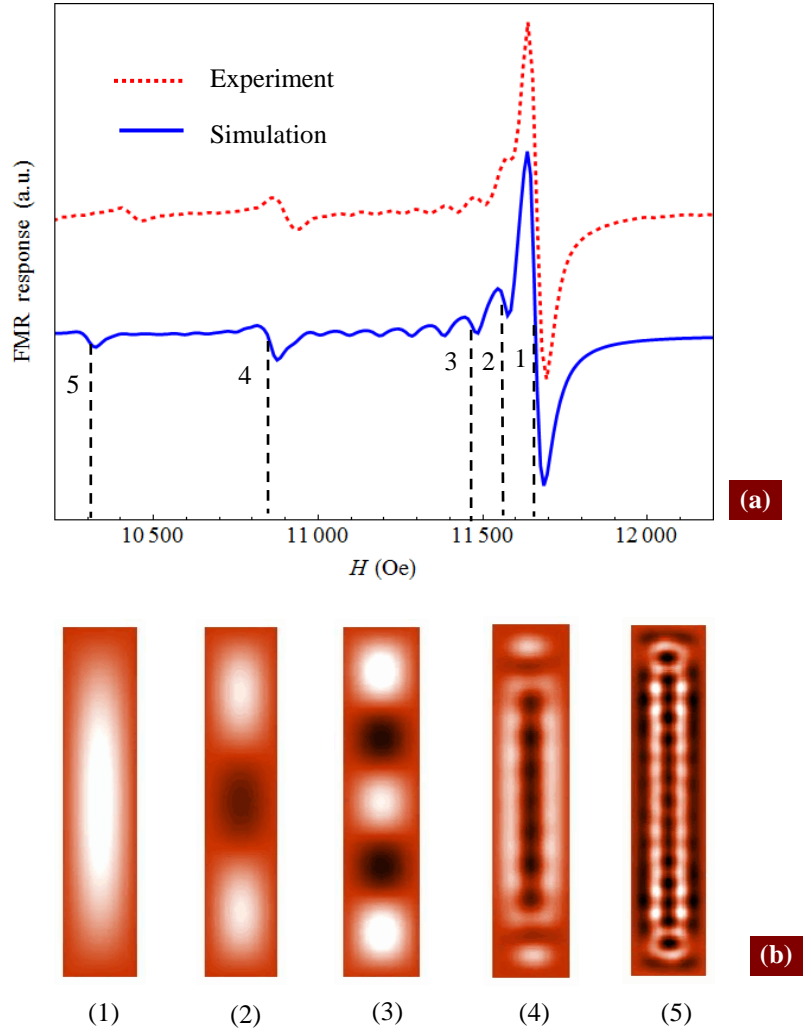

**Fig. 5.** (a) The experimental (dashed red curve) and simulated (solid blue curve) microstripe FMR (external magnetic field applied along Z axis). (b) the spatial distributions of the magnetization oscillations (amplitude of the z-component) in different resonant fields. (1) is for $H = 11660$ Oe; (2) is for $H = 11580$ Oe; (3) is for $H = 11480$ Oe; (4) is for $H = 10850$ Oe; (5) is for $H = 10320$ Oe.

**Conclusion**

Thus, the experimental results and micromagnetic modeling have shown that in the planar high aspect ratio microstripe the series of FMR resonances are observed. The structure of FMR spectra strongly depends on the sample orientation relative the magnetizing dc magnetic field. The simulated spatial distributions of resonant modes and the spin-wave spectra calculations shown that resonant oscillations have the complicated structure corresponding to the superposition of long-wave and short-wave modes, localized oscillations at the edges of microstrip. The transverse resonances are especially sensitive to the state of the strip edges (edge roughness, oxidation).

**Acknowledgements**

The authors are very thankful to A.A. Fraerman for the useful discussions and S.N. Vdovichev for the help in sample fabrication. We also thank M. J. Donahue and R. D. McMicheael (NIST, USA) for helpful councils. This work was supported by the Russian Foundation for Basic Research (project No. 15-02-04462)